\documentclass{aastex63}

\usepackage{graphicx}
\graphicspath{ {./Images/} }
\usepackage{amsmath}	
\usepackage{amssymb}

\submitjournal{APJ}

\shorttitle{Hot Super-Earths with Hydrogen Atmospheres}
\shortauthors{Modirrousta-Galian et al.}

\begin{document}

\title{Hot Super-Earths with Hydrogen Atmospheres: A Model Explaining Their Paradoxical Existence}

\correspondingauthor{Darius Modirrousta-Galian}
\email{darius.modirrousta@inaf.it}

\author[0000-0001-6425-9415]{Darius Modirrousta-Galian}
\affiliation{INAF – Osservatorio Astronomico di Palermo, Piazza del Parlamento 1, I-90134 Palermo, Italy}
\affiliation{University of Palermo, Department of Physics and Chemistry, Via Archirafi 36, Palermo, Italy}

\author{Daniele Locci}
\affiliation{INAF – Osservatorio Astronomico di Palermo, Piazza del Parlamento 1, I-90134 Palermo, Italy}

\author{Giovanna Tinetti}
\affiliation{Department of Physics \& Astronomy, University College London, Gower Street, WC1E 6BT London, United Kingdom}

\author{Giuseppina Micela}
\affiliation{INAF – Osservatorio Astronomico di Palermo, Piazza del Parlamento 1, I-90134 Palermo, Italy}

\begin{abstract}

In this paper we propose a new mechanism that could explain the survival of hydrogen atmospheres on some hot super-Earths. We argue that on close-orbiting tidally-locked super-Earths the tidal forces with the orbital and rotational centrifugal forces can partially confine the atmosphere on the nightside. Assuming a super terran body with an atmosphere dominated by volcanic species and a large hydrogen component, the heavier molecules can be shown to be confined within latitudes of $\lesssim 80^{\circ}$ whilst the volatile hydrogen is not. Because of this disparity the hydrogen has to slowly diffuse out into the dayside where XUV irradiation destroys it. For this mechanism to take effect it is necessary for the exoplanet to become tidally locked before losing the totality of its hydrogen envelop. Consequently, for super-Earths with this proposed configuration it is possible to solve the tidal-locking and mass-loss timescales in order to constrain their formation `birth' masses. Our model predicts that 55 Cancri e formed with a day-length between approximately $17-18.5$ hours and an initial mass less than $\rm \sim12 M_{\oplus}$ hence allowing it to become tidally locked before the complete destruction of its atmosphere. For comparison, CoRoT-7b, an exoplanet with very similar properties to 55 Cancri e but lacking an atmosphere, formed with a day-length significantly different from $\sim 20.5$ hours whilst also having an initial mass smaller than $\rm \sim9 M_{\oplus}$.

\end{abstract}



\section{Introduction} \label{sec:intro}

The composition of exoplanets and in particular super-Earths is a heavily disputed subject. The non-uniqueness of super-Earth interiors has lead to a plethora of geological models that could fit the observational data \citep[e.g.][]{Valencia2006,Valencia2007,Seager2007,Sotin2007,Rogers2010,Madhusudhan2012,Zeng2013,Zeng2016}. Fortunately, the recent introduction of atmospheric spectroscopy has helped to reduce the degeneracy of exoplanet compositions. However, the discovery of super-Earths like 55 Cancri e that have hydrogen-rich envelops \citep[e.g.][]{Tsiaras2016,Esteves2017} and effective temperatures above $\sim 2000$ $\rm K$ \citep{Demory2016} have raised new questions. This is because XUV atmospheric evaporation models predict the destruction of primordial atmospheres in million year timescales for planets with initial masses of super-Earths \citep[e.g.][]{Erkaev2007,Ehrenreich2011,Lammer2013,Owen2013,Jin2014,Owen2017,Kubyshkina2018(2),Kubyshkina2018(1),Locci2019}. Even when one considers the storage of primordial hydrogen into magma oceans which could in theory increase the total abundance by an order of magnitude \citep{Chachan2018}, mass-loss timescales would only be of the order $\sim$ few Myrs. Consequently, this apparent contradiction implies that there should be a mechanism which allows for hot super-Earth's with hydrogen atmospheres to exist.

In this manuscript we propose one potential solution to this conundrum for tidally locked low-mass low-density planets. The mechanism can be described by the following steps:
\begin{enumerate} 
	\item The super-Earth becomes tidally locked before losing the entirety of its primordial atmosphere. During this time active volcanism and surface vaporisation lead to the release of heavier gases which increase the mean molecular weight of the envelop.
	\item XUV irradiation unto the dayside triggers atmospheric losses which result in a pressure gradient across the hemispheres. Initially, the rate of atmospheric evaporation is balanced with the rate of advection meaning that mass-loss is rapid. In this `balanced phase' the XUV evaporation models present in the literature \citep[e.g.][]{Erkaev2007,Ehrenreich2011,Lammer2013,Owen2013,Jin2014,Owen2017,Kubyshkina2018(2),Kubyshkina2018(1),Locci2019} can be applied as mass-loss is limited by the XUV flux.
	\item The balanced phase lasts until the mean molecular mass of the atmosphere becomes large enough for the tidal and centrifugal forces to overcome the pressure gradient for the heavier gases. Hydrogen molecules are light enough to not be strongly influenced by the tidal and centrifugal forces so they are able to travel across the terminator and into the dayside. However, because the volcanic and mineral species are well confined to the nightside the hydrogen gas would need to diffuse through before escaping. This is analogous to Jean's escape on Earth which can only take place when light molecules reach the exobase after diffusing across the atmosphere. This process slows down the hydrogen-destruction rate considerably which we argue explains why some super-Earths are very old, hot and appear to host hydrogen-rich atmospheres. We label this chronological period as the `diffusive phase'.
\end{enumerate}

In this paper we apply our model to 55 Cancri e because of its well constrained mass and radius \citep{Crida2018}, its measured surface temperature distribution \citep{Demory2016} (which is also indicative of a synchronous rotation), and spectroscopic studies which are consistent with it being a terran body \citep{Ridden2016} hosting a hydrogen-rich atmosphere \citep{Tsiaras2016,Esteves2017}. CoRoT-7b being similar in mass and radius \citep{Barros2014} to 55 Cancri e but not being compatible with having an envelop \citep{Guenther2011,Kubyshkina2018(2)} has been used in order to act as a control planet. Furthermore, by solving the equations for the tidal locking timescale \citep{Gladman1996}, mass-loss timescale \citep{Erkaev2007,Kubyshkina2018(1)}, conservation of angular momentum \citep{Verbunt1993}, and the Roche limit \citep{Aggarwal1976}, the mass of a hot super-Earth at its birth (its initial mass) can be strongly constrained.

\section{A Simple Analytic Model}

In this section we will use 55 Cancri e's observed parameters which are shown in Table~(\ref{tab:55cnce}).
\begin{table}
	\centering
	\caption{The observed parameters of 55 Cancri e.}
	\label{tab:55cnce}
	\begin{tabular}{lccr} 
		\hline
		\hline
		Parameter & Symbol & Value & Reference \\
		\hline
		Radius ($\rm R_{\oplus}$) & $\rm R_{P}$ & $ 1.947 \pm 0.038 $ & (1) \\
		Mass ($\rm M_{\oplus}$) & $\rm M_{P}$  & $ 8.59 \pm 0.43 $ & (1) \\
		Semi-major axis (AU) & $\rm a_{P}$  & $\sim 0.016041 $ & (1) \\
		Period (days) & $\rm P_{P}$  & $\sim 0.7365474 $ & (1) \\
		Avg dayside temperature (K) & $\rm T_{day}$  & $ 2300^{+237}_{-144} $ & (2) \\
		Avg nightside temperature (K) & $\rm T_{night}$ & $ 1300^{+420}_{-375} $ & (2) \\		
		\hline
	\end{tabular}
\tablecomments{$\rm ^{1}$\citet{Crida2018}, $\rm ^{2}$\citet{Demory2016}}
\end{table}
The first step in modelling the nightside is to consider the transverse forces acting on the atmosphere. The dominant of these being the centrifugal force from the exoplanet's spin around its own axis ($\rm F_{spin}$), the centrifugal force from the orbit around the host star ($\rm F_{orb}$), the gravitational force from the host star ($\rm F_{g}$), and the gas pressure ($\rm F_{gas}$). Notwithstanding, these are not the only forces present as others such as atmospheric viscosity, atmosphere-surface friction, and the coriolis force would also take place. However, for the slow velocities predicted by our model these forces are negligible so they will therefore be ignored in order to keep in line with the simplicity of our analytic framework.

\subsection{Centrifugal Force from Spin}
\label{sec:spin} 

\begin{figure}
	\centering
	\includegraphics[scale = 0.75]{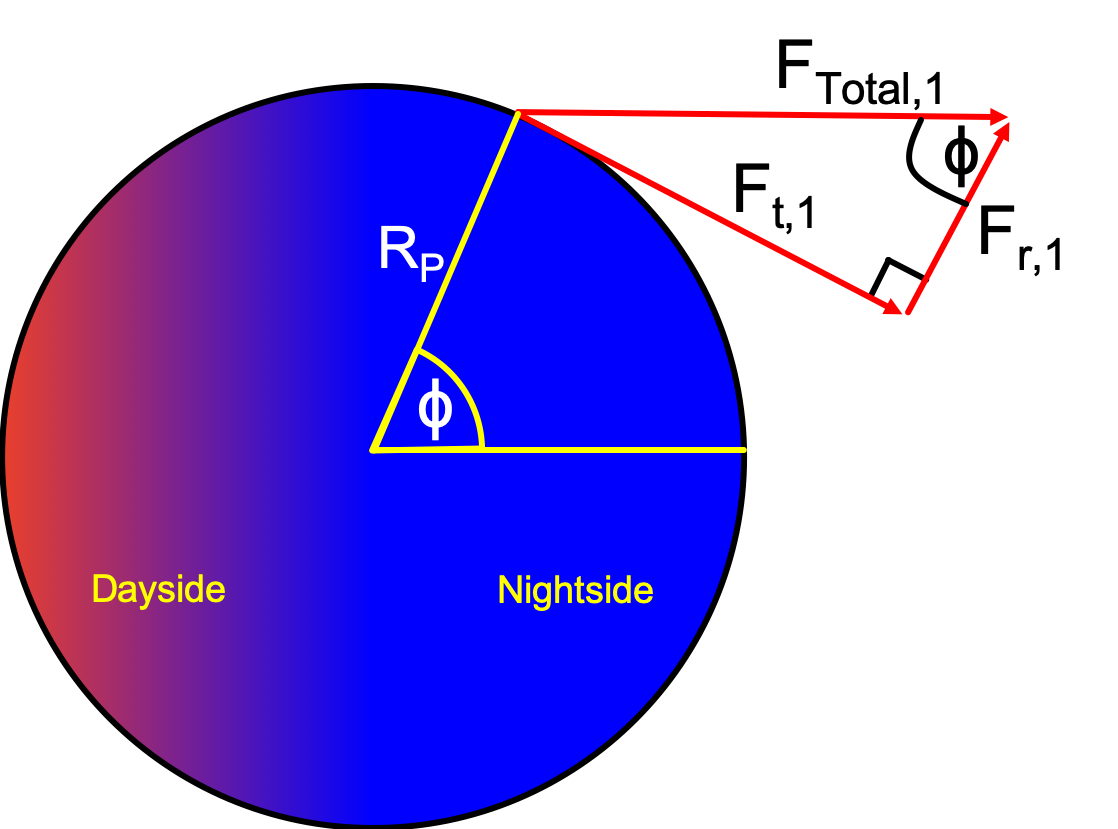}
	\caption{The different components of the centrifugal force where $F_{r,1}$ is the radial component, $F_{t,1}$ is the transverse component, and $F_{\rm Total,1}$ is the vector sum of the radial and transverse components. The angle $\phi$ is an arbitrary latitude on the nightside.}
	\label{fig:centrifugal}
\end{figure}
Using Fig~{\ref{fig:centrifugal}} as reference, the total centrifugal force , $F_{\rm Total,1}$, at an arbitrary latitude is
\begin{equation}
F_{\rm Total,1} = \omega^{2} R_{P} \cos{\left( \phi \right) } \times \delta m
\label{eq:spincent}
\end{equation}
Where $\delta m$ is an arbitrary unit of mass and $\omega$ is the angular velocity of the spin and orbit of 55 Cancri e which is calculated using the period ($P_{P}$) shown in Table~\ref{tab:55cnce}. However, if we want to find the transverse component we need to multiply by $\rm \sin{(\phi)}$ to get
\begin{equation}
F_{\rm spin} = F_{t,1} = \dfrac{1}{2}\omega^{2} R_{P} \sin{\left(2 \phi \right) }  \times \delta m
\label{eq:spincenttrans}
\end{equation}
This equation proposes that the maximum transverse component is at a latitude of $45^{\circ}$. At the limits, the centrifugal force from the planet's own spin is zero. This is because as $\phi \rightarrow 0^{\circ}$ the transverse component is nonexistent and when $\phi \rightarrow 90^{\circ}$ the total centrifugal force ($F_{\rm Total,1}$) vanishes.

\subsection{Centrifugal Force from Orbit}
\label{sec:orbit} 

In order to calculate this force we can also use Fig~{\ref{fig:centrifugal}} as a reference with the only difference being that the semi-major axis ($a_{P}$) also needs to be considered.
\begin{equation}
F_{\rm Total,2} = \omega^{2} \left(a_{P} + R_{P} \cos{\left( \phi \right) } \right)  \times \delta m
\label{eq:orbitcent}
\end{equation}
Similarly, if the transverse component is to be found, Equation~\ref{eq:orbitcent} needs to be multiplied by $\rm \sin{(\phi)}$ which gives
\begin{equation}
F_{\rm orb} = F_{t,2} = \omega^{2} \left(a_{P} + R_{P} \cos{\left( \phi \right) } \right)\sin{\left( \phi \right) }  \times \delta m
\label{eq:orbitcenttrans}
\end{equation}
Assuming $a_{P} \gg R_{P}$, the maximum force from the transverse component of the orbital centrifugal force occurs at $\phi = 90^{\circ}$ and the minimum is at $\phi = 0^{\circ}$.

\subsection{Stellar Gravitational Force}
\label{sec:stellar} 

\begin{figure}
	\centering
	\includegraphics[scale = 0.75]{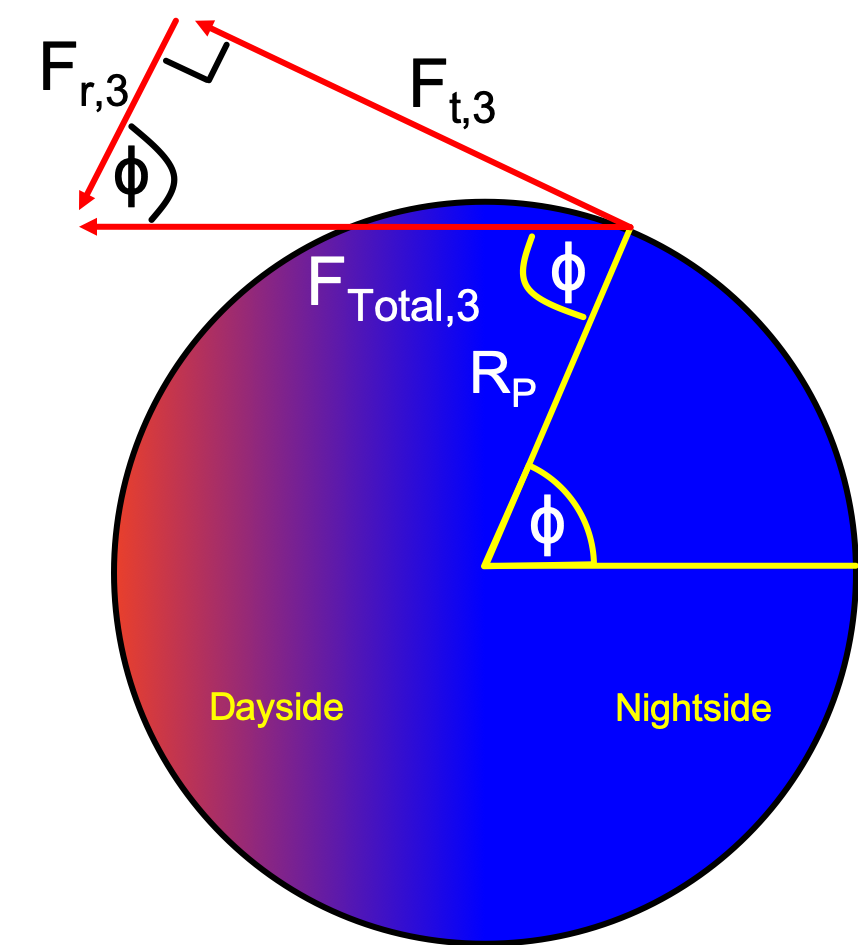}
	\caption{The different components of the stellar gravitational force where $F_{r,3}$ is the radial component, $F_{t,3}$ is the transverse component, and $F_{\rm Total,3}$ is the vector sum of the radial and transverse components. The angle $\phi$ is an arbitrary latitude on the nightside.}
	\label{fig:stellar_gravity}
\end{figure}
One important force which is often ignored is the star's gravity on the atmosphere. This force acts against the centrifugal forces and will pull the atmospheric gases towards the dayside. Fig~\ref{fig:stellar_gravity} shows the different components of the force from which Equation~\ref{eq:stellargrav} can be intuitively derived
\begin{equation}
F_{\rm Total,3} = \dfrac{GM_{\ast} \delta m}{\left(a_{P} + R_{P}\cos{\left( \phi\right) } \right)^{2} }
\label{eq:stellargrav}
\end{equation}
From this the transverse component can be found
\begin{equation}
F_{\rm g} = F_{t,3} = \dfrac{GM_{\ast} \delta m \sin{\left(\phi \right) }}{\left(a_{P} + R_{P}\cos{\left( \phi\right) } \right)^{2} }
\label{eq:stellargravtrans}
\end{equation}
According to Equation~\ref{eq:stellargravtrans} the maximum transverse force occurs at $\phi = 90^{\circ}$ and the minimum at $\phi = 0^{\circ}$. Concerning with the planet's own gravity, because this force would only act radially, we can ignore it. 

\subsection{Gas Force}
\label{sec:gas} 

\begin{figure}
	\centering
	\includegraphics[scale = 0.75]{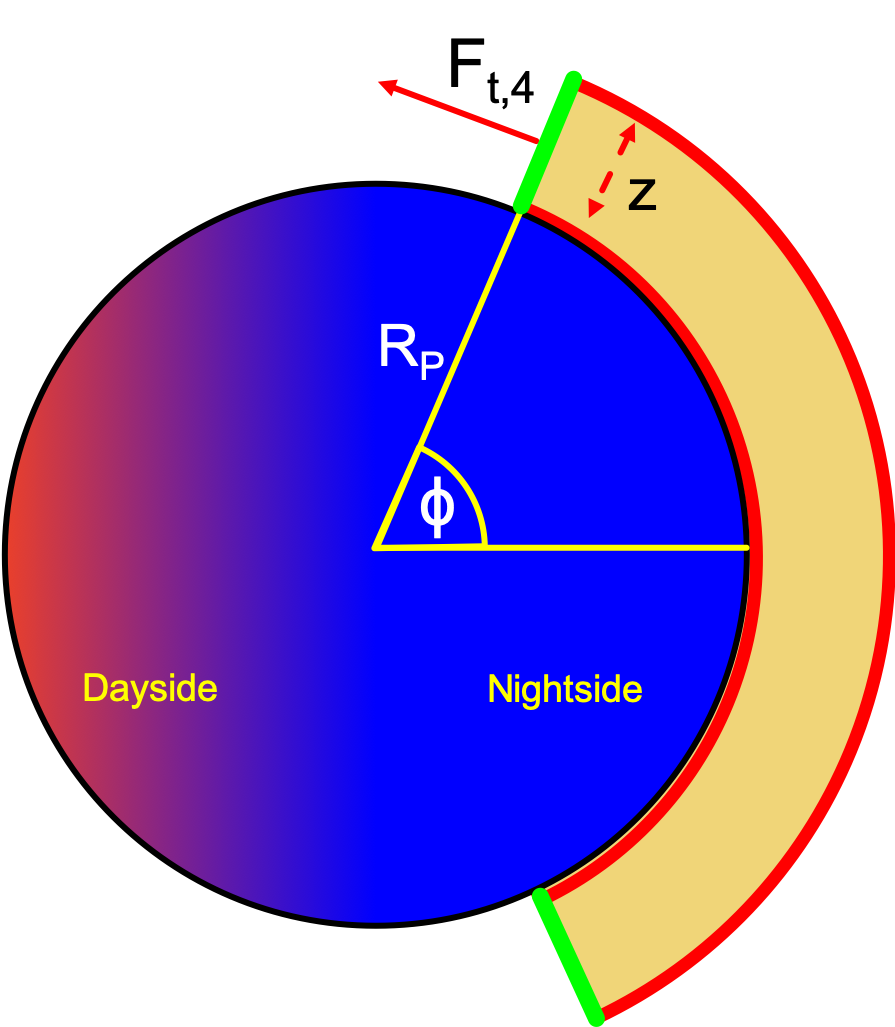}
	\caption{A simplified `ideal' representation of the resultant force from an atmosphere on the nightside of the planet. The green lines represent the cross-sectional area whilst the red lines show the total encased atmosphere within a given latitude. This diagram does not depict the configuration of 55 Cancri e, it is instead an extreme case scenario used to symbolise what the maximum gas force would be.}
	\label{fig:gasforce}
\end{figure}

The final force considered in this manuscript is the thermal gas force. Unsurprisingly, this is the hardest force to model, as many effects such as those related to the thermodynamics, viscous forces or turbulence should be considered. In order to avoid this complexity and analyse the situation analytically we can focus on the maximum possible gas force which occurs when there is a free expansion. Regarding temperature gradients, being on the nightside the temperatures should be relatively constant which is consistent with the phase curve data of 55 Cancri e \citep{Demory2016}. In our simple model we will therefore assume a constant nightside temperature. Furthermore, the high nightside temperatures inferred by \citet{Demory2016} mean that intermolecular forces become exceedingly small so it is acceptable to treat the gas ideally. Concerning with incoming thermal radiation, being tidally locked these rays should be partially blocked out. This can be shown by using an altered version of Equation 1 in \citet{Caldas2019} which gives the maximum extent that radiation from the host star can interact with the atmosphere on the nightside
\begin{equation}
\phi = \arcsin\left(\dfrac{R_{P}}{R_{P} + z} \right) 
\label{eq:atmlimit}
\end{equation}
A visual representation of this can be seen in figure 1 of \citet{Caldas2019}. In this equation, \textit{z} is the height of the atmosphere which is of the order $\sim \dfrac{k_{B}T_{\rm night}}{\bar{\mu} g_{P}}$ where $k_{B}$ is Boltzmann's constant, $\bar{\mu}$ is the mean molecular weight of the atmosphere and $g_{P}$ is the gravitational acceleration of the planet. To understand the amount of XUV irradiation that comes in contact with the nightside we first need to build our atmospheric model and then apply Equation~\ref{eq:atmlimit}. It is important to note that Figure~\ref{fig:gasforce} does not represent the configuration of 55 Cancrie, it just depicts the case where the gas force would be maximised. Instead Equation~\ref{eq:atmlimit} should be used in conjunction to Figure~\ref{fig:atmseparation} in order to visualise its applicability. We will therefore come back to this equation later and instead focus on modelling the gas force.
\begin{subequations}
	We begin with the ideal gas equation
	\begin{equation}
	P = \dfrac{\rho}{\bar{\mu}}k_{B}T_{\rm night}
	\end{equation}
	Where $\rho$ is the density of the atmosphere. To find the acceleration ($a$), we multiply by the cross-sectional area ($ A $) and divide by the encased atmospheric mass ($ M $). However, mass is volume multiplied by density so the formula can be simplied further.
	\begin{equation}
	a = \dfrac{k_{B}T_{\rm night}}{\bar{\mu}} \dfrac{A}{V}
	\end{equation}
	As shown in Fig~\ref{fig:gasforce} the area A is the green section which is approximately $A \approx 2 \pi R_{P} z \sin{\phi}$ and the volume is $V \approx 2 \pi R_{P}^{2} z \left(1- \cos{\phi} \right) $
	\begin{equation}
	a = \dfrac{k_{B}T_{\rm night}}{\bar{\mu} R_{P}} \cot{\left( \dfrac{\phi}{2}\right) }
	\end{equation}
	Therefore the gas force is
	\begin{equation}
	F_{gas} = F_{t,4} = \dfrac{k_{B}T_{\rm night}}{\bar{\mu} R_{P}} \cot{\left( \dfrac{\phi}{2}\right) } \times \delta m
	\label{eq:gasforce}
	\end{equation}
\end{subequations}
Equation~\ref{eq:gasforce} assumes that the atmosphere is expanding into a vacuum which is a strong simplification. Because of this it overestimates the gas force since a realistic atmosphere is continuous and does not have a discrete end. However, if we show that the centrifugal and gravitational forces can balance out with the maximum possible gas force then it follows that for more realistic situations the condition would also hold.

\section{The Atmospheric Model}

\subsection{Balanced Phase: Primordial Hydrogen Atmosphere}

\begin{figure}
	\centering
	\includegraphics[scale = 0.75]{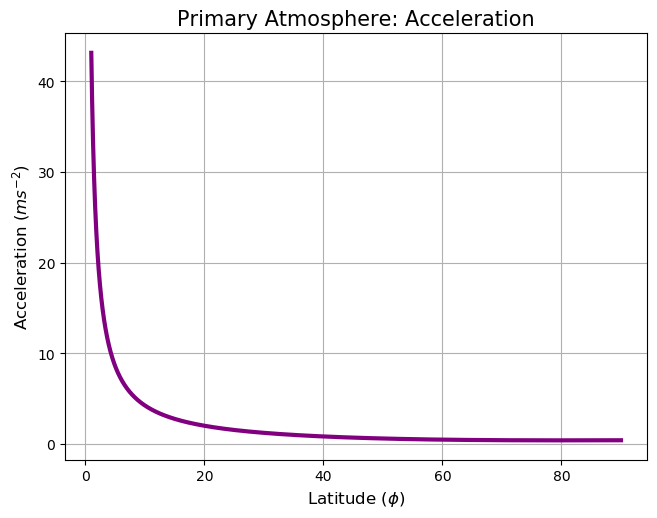}
	\caption{The total acceleration of the primordial atmosphere depending on its size which is measured as the latitude of maximum extent. Note how for all latitudes the acceleration is positive.}
	\label{fig:totalaccelerationprim}
\end{figure}

\begin{figure}
	\centering
	\includegraphics[scale = 0.75]{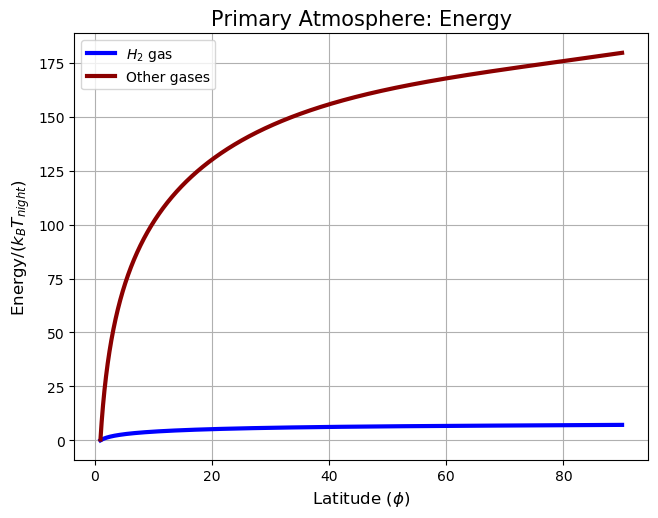}
	\caption{The energy of $\rm H_{2}$ (blue line) and volcanic molecules (red line) as a function of the latitude for the primordial atmosphere. The energy has been reduced by dividing through $k_{\rm B} T_{\rm night}$. }
	\label{fig:totalenergyprim}
\end{figure}

\begin{figure}
	\centering
	\includegraphics[scale = 0.75]{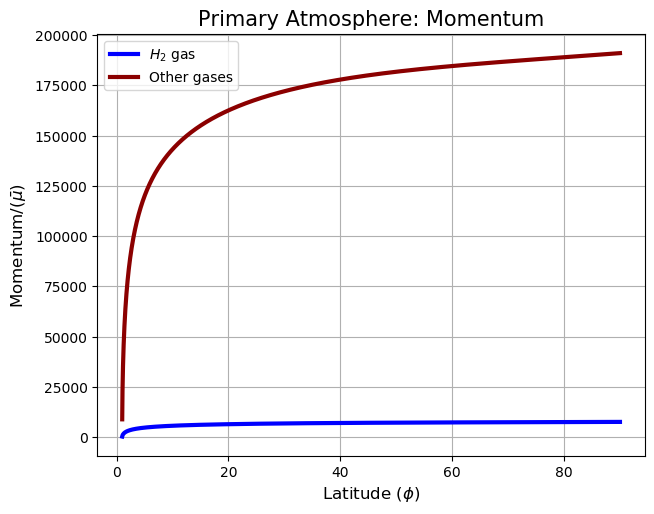}
	\caption{The momentum of $\rm H_{2}$ (blue line) and volcanic molecules (red line) as a function of the latitude for the primordial atmosphere. The momentum has been reduced by dividing through by the mean molecular mass of the atmosphere $\bar{\mu}$. }
	\label{fig:totalmomentumprim}
\end{figure}

We can first begin with a primordial atmosphere that has a mean molecular mass of $\bar{\mu}\approx 2.3$ amu if there is $> 99\%$ $\rm H_{2}$ ($\mu_{\rm H_{2}} \approx 2$ amu) and $< 1\%$ volcanic gases ($\mu_{\rm other} \approx 50$ amu). Since the atmosphere is dominated by hydrogen we can ignore the chemical reactions with the volcanic gases and the subsequent products. If we combine Equations~\ref{eq:spincenttrans},\ref{eq:orbitcenttrans},\ref{eq:stellargravtrans} and \ref{eq:gasforce} and divide by the arbitrary mass unit $\delta m$ we get the total acceleration as a function of the latitude which is shown in Fig~\ref{fig:totalaccelerationprim}.
To calculate the kinetic energy and momentum of the gases we follow a simple algorithm that considers the acceleration at each latitude for each chemical species. The results are shown in Fig~\ref{fig:totalenergyprim} and Fig~\ref{fig:totalmomentumprim}.

The graphs show that when the atmosphere is dominated by hydrogen there is a positive force which causes the gases to advect towards the dayside where they would be dissociated by the incoming XUV irradiation. The interesting outcome is that since all the gases are accelerated equally, the heavier volcanic and mineral species have greater absolute energies because of their larger masses. Our calculations show that for a primordial atmosphere the gas pressure would overcome the tidal and centrifugal forces so gas would be able to flow approximately freely into the dayside. In other words, atmospheric destruction would be balanced out by the mass flux from the nightside. The balanced phase would take place until the loss of hydrogen, and active volcanism would raise the mean molecular weight of the atmosphere to a level where tidal and centrifugal forces became important. Our model predicts that for a planet like 55 Cancri e the critical mean molecular weight is at $\bar{\mu} \approx 35$ amu ($\sim 70\%$ volcanic gases and $\sim 30\%$ $\rm H_{2}$) beyond which the tidal forces dominate over the maximum gas force.

\subsection{Diffusive Phase: Volcanic Atmosphere}
\label{sec:diff}

\begin{figure}
	\centering
	\includegraphics[scale = 0.75]{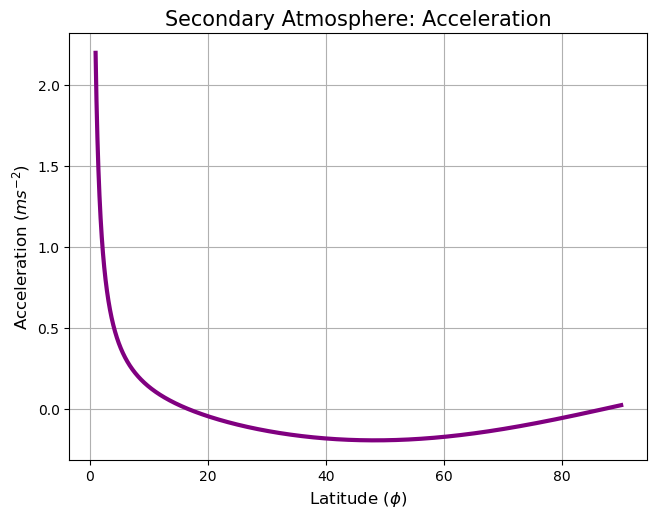}
	\caption{The total acceleration of the evolved heavier atmosphere depending on its size which is measured as the latitude of maximum extent. Note how for latitudes greater than $\sim 15^{\circ}$ and smaller than $\sim 85^{\circ}$ the acceleration is negative. }
	\label{fig:totalaccelerationvolc}
\end{figure}

\begin{figure}
	\centering
	\includegraphics[scale = 0.75]{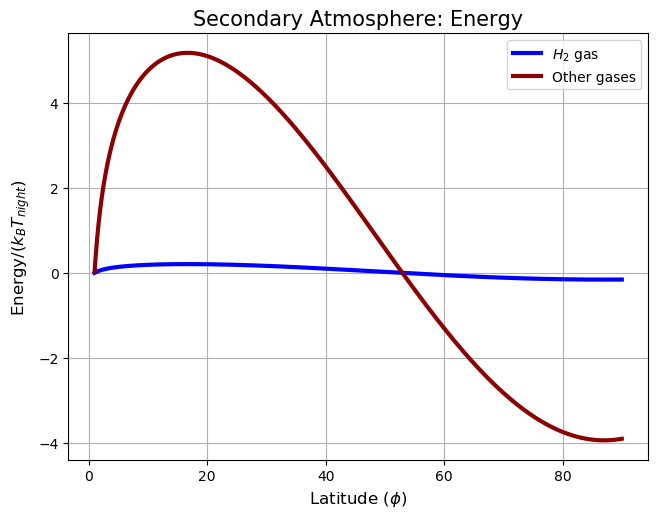}
	\caption{The energy of $\rm H_{2}$ (blue line) and volcanic molecules (red line) as a function of the latitude for the evolved heavier atmosphere. Note how for latitudes greater than $\sim 55^{\circ}$ the energies of the molecules are negative. The energy has been reduced by dividing through $k_{\rm B} T_{\rm night}$. }
	\label{fig:totalenergyvolc}
\end{figure}

\begin{figure}
	\centering
	\includegraphics[scale = 0.75]{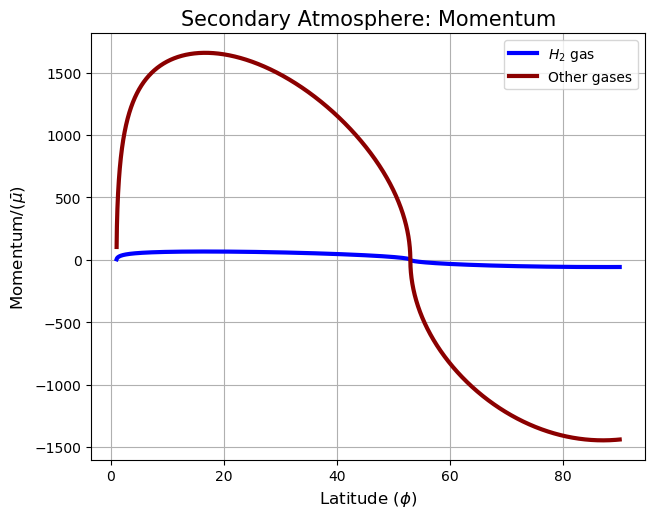}
	\caption{The momentum of $\rm H_{2}$ (blue line) and volcanic molecules (red line) as a function of the latitude for the evolved heavier atmosphere. Note how for latitudes greater than $\sim 55^{\circ}$ the momentum of the molecules is negative. The momentum has been reduced by dividing through by the mean molecular mass of the atmosphere $\bar{\mu}$. }
	\label{fig:totalmomentumvolc}
\end{figure}

\begin{figure}
	\centering
	\includegraphics[scale = 0.65]{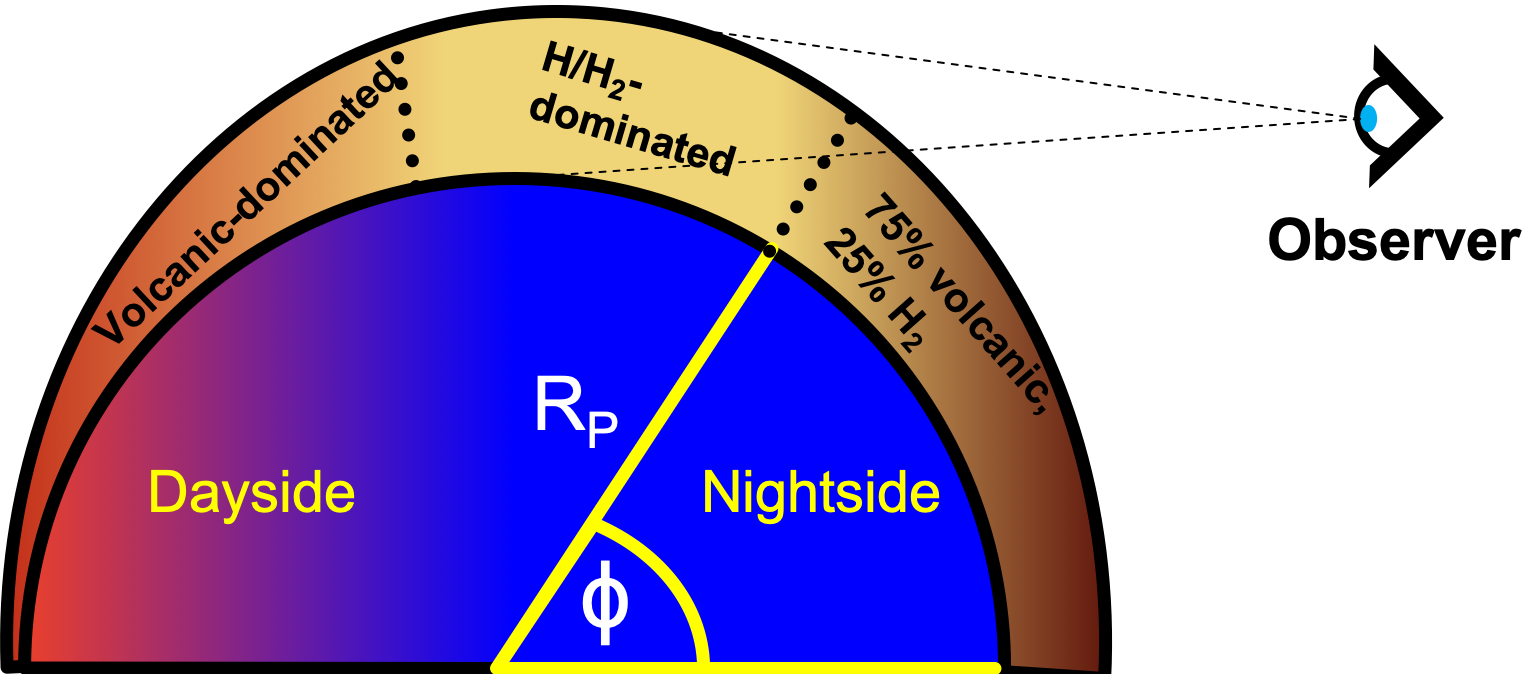}
	\caption{A simplified illustration of the atmospheric distribution predicted on some hot super-Earths with hydrogen atmospheres. This image does not illustrate the atmospheric density at the different latitudes.}
	\label{fig:atmseparation}
\end{figure}

Once the planet reaches the diffusive phase the atmospheric dynamics would change. To demonstrate this phenomenon we assume that the nightside atmosphere is $\sim 10\%$ hydrogen and $\sim 90\%$ volcanic and mineral species which gives $\bar{\mu}\approx 45$ amu. Clearly a full chemical analysis would be required to account for the interactions between carbon and hydrogen which have a tendency to react. However, as a proof of concept we will work with these values. Our results for the acceleration, energy and momentum are shown in Fig~\ref{fig:totalaccelerationvolc}, \ref{fig:totalenergyvolc} and \ref{fig:totalmomentumvolc} respectively.
For an atmosphere that extends to $ \phi \approx 80^{\circ} $ the volcanic species would need $\sim 4k_{\rm B}T_{\rm night} $ to escape whilst the $\rm H_{2} $ molecules would need $ \sim 0.1 k_{\rm B}T_{\rm night} $. It is important to note that if the atmospheric viscosity, friction, turbulance, and a less extreme gas force were used then the required escape energies would be greater. Notwithstanding, due to the Maxwell-Boltzmann distribution of velocities it would be easier for the $\rm H_{2} $ molecules to escape to the dayside than the heavier species. However, since the volcanic and mineral gases are more hydrodynamically stable due requiring more energy to escape, the $\rm  H_2 $ molecules would not be able to flow unobstructed into the dayside. Instead, the hydrogen gas would need to diffuse through the atmosphere in a similar fashion to how hydrogen molecules on Earth can only escape when they reach the exosphere. This diffusion will slow down the $\rm H_2 $ loss considerably which can be shown by applying Ficks $1^{\rm st}$ law of diffusion.
\begin{equation}
\dfrac{dM}{dt} = \dfrac{\Delta \rho_{H_{2}} \cdotp A \cdotp D}{L}
\label{eq:ficks1st}
\end{equation}
Where $\Delta \rho_{H_{2}}$ is the hydrogen density contrast between both hemispheres, $A$ is the cross-sectional area of the atmosphere, $D$ is the diffusion constant, and $L$ is the transverse lengthscale of the diffusive motions. Calculating the diffusive flux is a complex dynamic problem but using Equation~\ref{eq:ficks1st} with a few `guesstimates' for the parameters will show how in principle the mass flux would be greatly suppressed. Using $\Delta \rho_{H_{2}} \sim 1$ $\rm kg$ $\rm m^{-3}$ (unimportant because $ D\propto \rho_{H_{2}}^{-1} $), $A \sim 2 \pi R_{P} z$, $L\sim R_{P}$, and $D \sim 10^{-5} $ $\rm m^{2}$ $\rm s^{-1}$ (calculated using Chapman-Enskog theory) gives $\rm \dfrac{dM}{dt} \sim 1$ $\rm kg$ $\rm s^{-1}$. The uncertainty in this value could be of several orders of magnitude especially since the diffusion constant can vary greatly depending on the turbulence present, but the important point to take from this calculation is that diffusive transport is slower than a free expansion. Notwithstanding, due to there being more hydrogen molecules traversing into the dayside than volcanic and mineral species, the terminator region would become hydrogen-rich as illustrated in Fig~\ref{fig:atmseparation}. The hydrogen gas would then be removed from the planet due to XUV irradiation \citep[e.g.][]{Erkaev2007,Ehrenreich2011,Lammer2013,Owen2013,Jin2014,Owen2017,Kubyshkina2018(2),Kubyshkina2018(1),Locci2019}.
This is consistent with spectroscopic studies of the terminator which have found evidence of a hydrogen-rich envelop on 55 Cancri e \citep[e.g.][]{Tsiaras2016,Esteves2017} despite being old and hot. Furthermore, this is also compatible with observations that were unable to detect significant hydrogen destruction in the exosphere of 55 Cancri e \citep{Ehrenreich2012, Bourrier2018(1)}.

Now we can come back to Equation~\ref{eq:atmlimit} which describes the maximum latitude within which stellar light can interact with the nightside's atmosphere. In the diffusive phase the mean molecular mass will be large ($\gtrsim 35$ amu) so the atmospheric scale height will be small. Plugging reasonable values into Equation~\ref{eq:atmlimit} shows that latitudes $\lesssim 85^{\circ}$ would be safe from XUV irradiation. This is consistent with our previous assumption that for an atmosphere that extends to $ \phi \approx 80^{\circ}$, thermal fluctuations can be ignored.

\section{Tidal Locking and Atmospheric Loss}
\label{sec:tidal}

\begin{figure}[h]
	\centering
	\includegraphics[scale = 0.75]{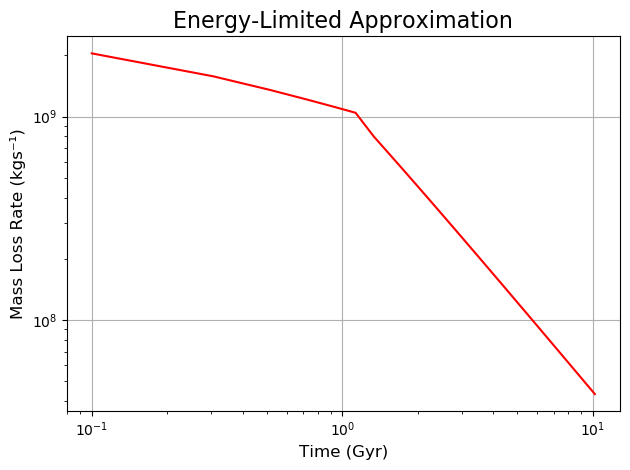}
	\includegraphics[scale = 0.75]{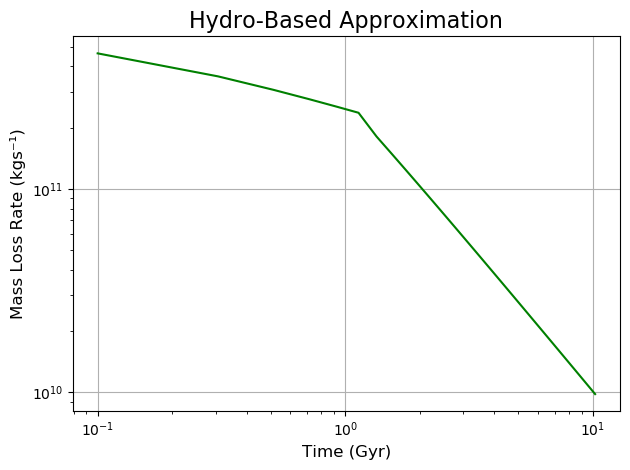}
	\caption{A side-by-side comparison of the mass-loss rates predicted by the energy-limited approximation \citep{Erkaev2007} and the hydro-based approximation \citep{Kubyshkina2018(1)}. Note the different scale on the y-axis.}
	\label{fig:comparison}
	\centering
\end{figure}

\begin{figure}[h]
	\centering
	\includegraphics[scale = 0.75]{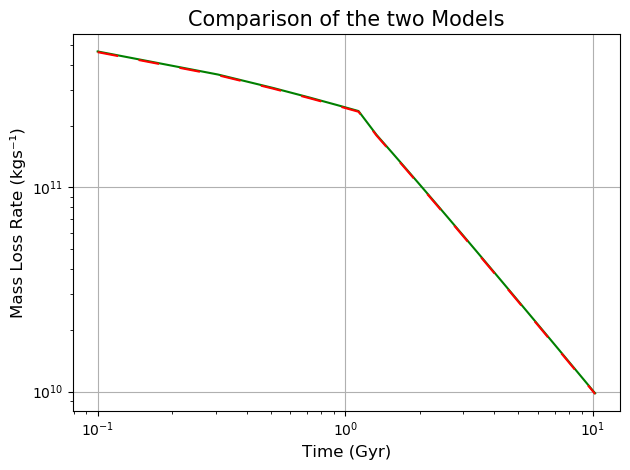}
	\caption{A comparison of the mass-loss rates predicted by the corrected energy-limited approximation \citep{Erkaev2007} and the hydro-based approximation \citep{Kubyshkina2018(1)}. The green line marks the hydro-based approximation while the red dashed line represents the corrected energy-limited approximation.}
	\centering
	\label{fig:225correction}
\end{figure}

As explained previously, for our model to take effect it is essential that the super-Earth in question becomes tidally locked before losing the totality of its primordial atmosphere. Prior to this time the entire atmosphere was exposed to strong X-Ray and UV radiation so mass-loss was global. It can therefore be inferred that the further back in time one travels, the larger and more massive the hydrogen envelop was. The first step is therefore to calculate when the tidal locking occurred which can be approximated by:

\begin{subequations}	
	\begin{equation}
	\ t_{\rm lock} \approx \dfrac{\omega_{0} a^{6}_{0} I_{0} Q_{0}}{3 G M^{2}_{0} k_{2} R^{5}_{0}}
	\label{eq:tidallocking}
	\end{equation}
	where
	\begin{equation}
	\omega_{0} = |\omega_{\rm orb} - \Delta \omega| 
	\end{equation}  	
\end{subequations}
$ \omega_{0} $ is the initial spin rate relative to the orbital period of the planet-star system ($\rm rad $ $\rm s^{-1} $), $ \omega_{\rm orb} $ is the current spin rate required for the planet to be tidally locked, $ \Delta \omega $ is the initial intrinsic spin rate of the planet, $ a_{0} $ is the initial semi-major axial distance of the planet ($\rm m $), $ I_{0} $ is the initial moment of inertia of the planet ($\rm kg$ $\rm m^{2}$), $ Q_{0} $ is the initial dissipation function of the planet (dimensionless), $ M_{0} $ is the initial mass of the planet ($\rm kg $), $ k_{2} $ is the initial tidal love number of the satellite (dimensionless) and $ R_{0} $ is the initial radius of the planet ($\rm m $) \citep{Gladman1996}.There are two main issues that arise when trying to use Equation~\ref{eq:tidallocking}:
\begin{enumerate}
	\item The initial spin rate ($ \omega_{0}$) is completely unknown and cannot be predicted due to its
	inherent chaotic nature \citep{Miguel2010}.
	\item $ a_{0} $, $ I_{0} $, $ Q_{0} $ and $ k_{2} $ are heavily dependent on $ M_{0} $ or $ R_{0} $ or both, which themselves must be calculated from X-Ray and UV mass-loss models.
\end{enumerate}
Consequently, the tidal-locking and mass-loss timescales must be solved simultaneously. Before this can be done, it is necessary to select the appropriate mass-loss model. On the one hand, whilst the energy-limited formula \citep{Erkaev2007} is simple and has a similar behaviour, it underestimates the true value by two orders of magnitude for super-Earths and sub-Neptunes (see Figure~\ref{fig:comparison}). On the other hand, the hydro-based approximation \citep{Kubyshkina2018(1)} is far more complex but accurately follows the trend and mass-loss for super-Earths.
Since we are only dealing with super-Earths and sub-Neptunes, a fudge factor of 225 can be applied to the energy-limited approximation to make it perfectly agree with the hydro-based model (see Figure~\ref{fig:225correction}), whilst also maintaining its simplicity.
\begin{equation}
\ \dfrac{dM}{dt} = 225 \frac{\varepsilon \pi R^{3}_{0} F_{XUV}}{GM_{0}K}
\label{eq:fixedmassloss}
\end{equation}
The corrected energy limited equation where $\varepsilon$ is the escape efficiency, $F_{\rm XUV}$ is the X-Ray and UV radiation flux ($\rm J$ $\rm m^{-2}$ $\rm s^{-1}$) \citep{Penz2008(2),Sanz2011}, and $K$ is the energy reduction factor which accounts for stellar tidal forces \citep{Erkaev2007}. With this information the mass-loss timescale can be approximated:
\begin{equation}
\ t_{\rm mass \: loss} \approx \dfrac{M_{0} - M_{P}}{\dot{M}} = \dfrac{G K M_{0} (M_{0} - M_{P})}{225 \varepsilon \pi R^{3}_{0} F_{\rm XUV}}
\label{eq:masslosstimescale}
\end{equation}
Parameters with a subscript $ 0 $ are for the original exoplanet before it experienced mass-loss, while the parameters with a subscript $ P $ are for the current exoplanet. Finally, the last effect that needs to be considered is planetary migration due to the conservation of angular momentum. For a planetary system were $ \sim50\% $ of all the angular momentum is conserved, the necessary condition is given by Equation~\ref{eq:verbunt} \citep{Verbunt1993}:
\begin{equation}
\ a_{0} \approx  a_{P}\left(\dfrac{M_{P}}{M_{0}} \right)^{7/4} 
\label{eq:verbunt}
\end{equation}
The conservation of angular momentum gives the relationship between the initial semi-major axis $ a_{0} $ ($\rm m $) with the initial mass $ M_{0} $ ($\rm kg $), the final mass $ M_{P} $ ($\rm kg $), and the final semi-major axis $ a_{P} $ ($\rm m $). As shown before, if a hot exoplanet becomes tidally locked before losing all of its atmosphere, it might retain a portion of its original hydrogen envelop. Therefore, this occurs if condition \ref{eq:minmasscondition} is met; which in practice means that the presence of a hydrogen-rich atmosphere on a super-Earth or sub-Neptune depends on the initial properties of the system. This can be mathematically modelled to estimate the minimum initial mass of an exoplanet:
\begin{subequations}
	\begin{equation}
	t_{\rm mass\:loss} > t_{\rm lock}
	\label{eq:minmasscondition}
	\end{equation}
	Combining Equations~\ref{eq:tidallocking}, \ref{eq:masslosstimescale}, and \ref{eq:verbunt} with some algebra (as shown in the appendix) gives
	\begin{equation}
	M_{0} \gtrsim \left( \alpha \cdot \left|  \dfrac{1}{T_{\rm orb}} - \dfrac{1}{\Delta T} \right| ^{1/10} \cdot M^{7/10}_{P} \cdot a^{2/5}_{P} \right)  + \dfrac{M_{P}}{10} 
	\label{eq:minmassderive}
	\end{equation}
	\label{eq:minmasstotal}
\end{subequations}

We obtain the minimum initial mass of a hot super-Earth retaining a hydrogen-rich atmosphere where $ \alpha =\left( \dfrac{45 \pi \varepsilon I_{XUV} Q_{0}}{4 G^{2} K k_{2}} \right)^{1/10} $. All the parameters are the same as those shown used in previous equations except for $ T_{\rm orb}$ and $ \Delta T$ which are the current orbital period and the initial intrinsic day-length respectively. One unexpected outcome of Equation~\ref{eq:minmassderive} is that calculating constant $ \alpha$ precisely is not a problem since being raised to the $\rm 1/10^{th}$ power greatly supresses uncertainties such that only the order of magnitude is important. Consequently, we used the values for the parameters presented in table~\ref{tab:alpha}.
\begin{table}
	\caption{The first order of magnitude estimates for the initial conditions of 55 Cancri e}
	\label{tab:alpha}
	\centering
	\begin{tabular}{p{1.5cm}p{2cm}p{4cm}} 
		\hline
		\hline
		Parameter & Value & Reference \\
		\hline                        
		$ K$ & $\sim 1$ & \citet{Erkaev2007} \\
		$ k_{2}$ & $\sim 10^{-1}$ & \citet{Driscoll2015} \\
		$ \varepsilon$ & $\sim 10^{-1}$ & \citet{Locci2019} \\
		$ I_{\rm XUV}$ & $\sim 10^{23}$ ($\rm W$) & \citet{Penz2008(2),Sanz2011} \\
		$ Q_{0}$ & $\sim 10^{2}$ & \citet{Jackson2008,Driscoll2015,Clausen2015} \\
		\hline 
	\end{tabular}
\end{table}
Using these values gives $ \alpha \approx 5 \times 10^{4}$. Substituting this into Equation~\ref{eq:minmassderive} gives the final formula for the minimum initial mass required for a hot super-Earth to have hydrogen in its atmosphere:
\begin{equation}
M_{0} \gtrsim \left(5\times 10^{4} \cdot \left|  \dfrac{1}{T_{\rm orb}} - \dfrac{1}{\Delta T} \right|^{1/10} \cdot M^{7/10}_{P} \cdot a^{2/5}_{P} \right)  + \dfrac{M_{P}}{10}
\label{eq:minmass}
\end{equation}

\subsection{The Maximum Initial Mass}

One important aspect that has been readily ignored in previous studies of exoplanet evaporation is the migration induced due to the conservation of angular momentum; as an exoplanet loses mass it must migrate outwards. However, due to the Roche limit \citep{Aggarwal1976}, there is a bound to how close an exoplanet could be to its host star in the past.
\begin{equation}
\ a_{\rm RL} \approx \mu R_{\ast} \left(\dfrac{\rho_{\ast}}{\rho_{0}} \right)^{1/3}
\label{eq:rochelimit} 
\end{equation}
Equation~\ref{eq:rochelimit} is the Roche limit where $ \mu$ is a constant of value $ 1.957 $, $ R_{\ast} $ is the radius of the host star (m), $ \rho_{\ast} $ is the density of the host star ($\rm kg $ $\rm m^{-3} $), and $ \rho_{0} $ is the initial density of the planet ($\rm kg $ $\rm m^{-3} $) \citep{Aggarwal1976}. Therefore, combining the conservation of angular momentum (Equation~\ref{eq:verbunt}) with the Roche limit (Equation~\ref{eq:rochelimit}) one can arrive at the theoretically maximum initial mass that a super-Earth could have had:
\begin{subequations}	
	\begin{equation}
	a_{0} > a_{\rm RL}
	\end{equation}
	Substituting in Equation~\ref{eq:verbunt} and \ref{eq:rochelimit}
	\begin{equation}
	\left(\dfrac{M_{P}}{M_{0}} \right)^{7/4} a_{P} > \mu R_{\ast} \left(\dfrac{\rho_{\ast}}{\rho_{0}} \right)^{1/3}
	\end{equation}	
	For hot exoplanets with large hydrogen envelops $ \rho_{0} \sim \rho_{\ast}$
	\begin{equation}
	\left(\dfrac{M_{P}}{M_{0}} \right)^{7/4} a_{P} \gtrsim \mu R_{\ast}
	\end{equation}
	Using trivial algebra
	\begin{equation}
	M_{0} \lesssim \beta \cdot M_{P} \cdot a^{4/7}_{P}
	\label{eq:maxmasstotal}
	\end{equation}
	\label{eq:maxmassderive}	
\end{subequations}
where $ \beta =\left( \mu R_{\ast} \right)^{-4/7} $. By using $ R_{\ast} \approx 6.8 \times 10^{8} $ $\rm m $ \citep{Crida2018} and $ \mu \approx 1.957 $ \citep{Aggarwal1976} we calculate the constant to be $ \beta \approx 6 \times 10^{-6} $. Substituting this into Equation~\ref{eq:maxmasstotal} gives the formula for the maximum initial mass physically possible:
\begin{equation}
\ M_{0} \lesssim 6 \times 10^{-6} \cdotp M_{P} \cdotp a^{4/7}_{P}
\label{eq:maxmass}
\end{equation}

\section{Results}

\begin{figure}[h]
	\centering
	\includegraphics[scale = 0.75]{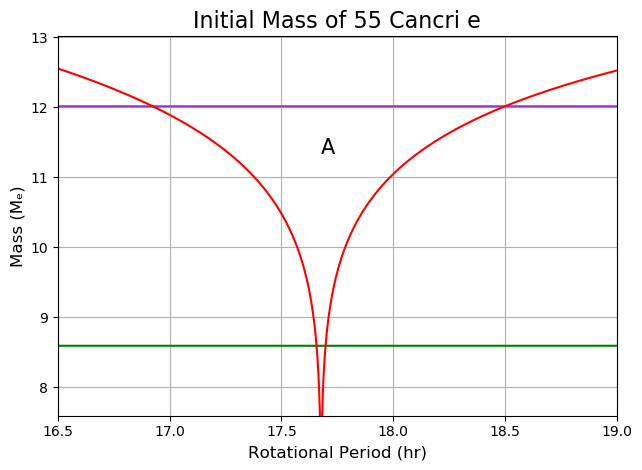}
	\caption{The initial mass of 55 Cancri e. The enclosed region marked with the ‘A’ shows the permitted ranges for the initial mass. The red line shows the initial mass threshold below which 55 Cancri e would not have a hydrogen-rich atmosphere. The purple line marks the maximum initial mass of 55 Cancri e due to the Roche limit \citep{Aggarwal1976}, and the green line is the current mass of 55 Cancri e.}
	\centering
	\label{fig:55cancrie}
\end{figure}

With the minimum and maximum limits of the initial mass known, it is now possible to show the mass range of 55 Cancri e when it formed. Figure~\ref{fig:55cancrie} shows the permitted ranges for the initial mass (the enclosed region marked with the ‘A’).The red line shows the initial mass threshold below which 55 Cancri e would not have a hydrogen-rich atmosphere. The purple line marks the maximum initial mass of 55 Cancri e due to the Roche limit \citep{Aggarwal1976}, and the green line is the current mass of 55 Cancri e.

CoRoT-7b is a super-Earth with similar properties to 55 Cancri e as shown by table~\ref{tab:corot7bproperties}. In spite of experiencing strong tidal and centrifugal forces, CoRoT-7b most likely lacks a hydrogen atmosphere \citep{Kubyshkina2018(2),Guenther2011}. According to our model, this is due to its original mass and day-length being below the necessary threshold for the tidal-locking timescale to be shorter than the complete mass-loss timescale. We argue that at one point in its past Corot-7b had a hydrogen atmosphere, but the asynchronous rotational period meant that the atmosphere was exposed to harmful XUV radiation which resulted in its eventual depletion.
\begin{table}
	\caption{The observed parameters of CoRoT-7b}
	\label{tab:corot7bproperties}
	\centering
	\begin{tabular}{p{3.5cm}p{2cm}p{2cm}} 
		\hline
		\hline
		Parameter & Value\\
		\hline                  
		Mass ($\rm M_{\oplus}$) & $5.74\pm0.86$$ ^{1} $\\
		Radius ($\rm R_{\oplus}$) & $1.585\pm0.064$$ ^{1} $\\
		Density ($\rm \rho_{\oplus}$) & $1.19\pm0.27$$ ^{1} $\\
		Mean temperature ($\rm K $) & $\sim 1800$\\
		Semi-major axis ($\rm AU $) & $0.017016$$ ^{1}$\\
		Host star spectal type & G9 V\\
		\hline
	\end{tabular}
\tablecomments{$ ^{1} $\citet{Barros2014}}
\end{table}
\begin{figure}[h]
	\centering
	\includegraphics[scale = 0.75]{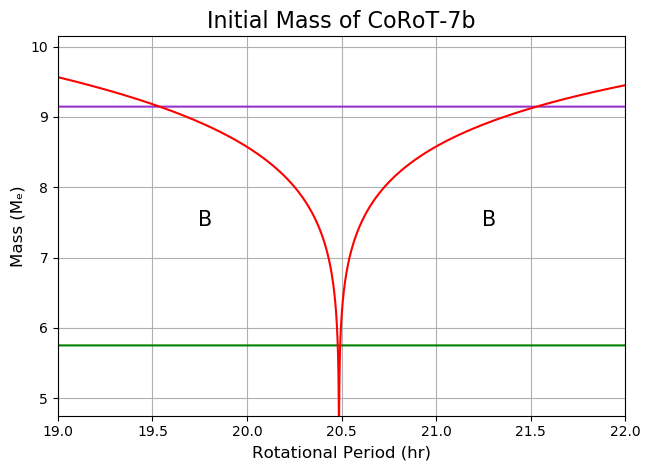}
	\caption{The initial mass of CoRoT-7b. The enclosed region marked with the ‘B’ shows the permitted ranges for the initial mass. The red line shows the initial mass threshold beyond which CoRoT-7b would host a hydrogen-rich atmosphere, the purple line marks the maximum initial mass of CoRoT-7b due to the Roche limit \citep{Aggarwal1976}, and the green line is the current mass of CoRoT-7b.}
	\centering
	\label{fig:corot7b}
\end{figure}

On figure~\ref{fig:corot7b} the enclosed region marked with the ‘B’ shows the permitted ranges for the initial mass. The red line shows the initial mass threshold beyond which CoRoT-7b would host a hydrogen atmosphere, the purple line marks the maximum initial mass of CoRoT-7b due to the Roche limit \citep{Aggarwal1976}, and the green line is the current mass of CoRoT-7b.

\section{Discussions}

Our model indicates that 55 Cancri e could host a metastable atmosphere with a substancial hydrogen component on its nightside which is out of reach from the host star’s UV and X-Ray radiation that, unobstructed, would have caused its destruction. We argue this configuration was achieved because 55 Cancri e became tidally locked before losing the totality of its hydrogen depository. This requires the initial rotational period to be between approximately $17-18.5$ hours as well as the initial mass being smaller than $\rm \sim12 M_{\oplus} $. In contrast, CoRoT-7b most likely lost all of its atmosphere before becoming tidally locked which resulted in it lacking a hydrogen envelop. Knowing this, we calculated that CoRoT-7b must have formed with an initial rotational period significantly different from $\sim 20.5$ hours as well as its initial mass being smaller than $\rm  \sim 9 M_{\oplus} $.

\subsection{Assumptions Used in Our Model}

To reach this conclusion we made several assumptions. In the following we discuss their implications and if and how they limit the use of our model to super-Earths with high temperatures orbiting G-, K-, M-type stars:
\begin{enumerate}
	\item\textit{In this paper we assume that 55 Cancri e has an atmosphere with a considerable hydrogen component.} The geology of 55 Cancri e is an active subject of dispute although the more precise mass and radius measurements from \citet{Crida2018} have lowered the degeneracy considerably. Using the geological models from \citet{Zeng2013,Zeng2016} and considering the limits permitted by the uncertainty in \citet{Crida2018}, it can be shown that the mass and radius of 55 Cancri e is not consistent with a denuded terrestrial planet even if it is coreless. However, other studies that measured the mass and radius of 55 Cancri e \citep[e.g.][]{Bourrier2018(2)} claim values that could be consistent with a coreless terrestrial planet, although they state it is unlikely. In addition, having no core introduces new problems such as the planet lacking a magnetic field which negates the possibility of magnetic planet-star interactions \citep[e.g.][]{Winn2011} that could explain the observed  $\sim 100$ $\rm ppm$ modulation in the optical \citep{Winn2011,Dragomir2014} and IR \citep{Tamburo2018} phase curves. Other proposed models for 55 Cancri e include the possibility of it being an icy planet \citep{Zeng2013,Zeng2016} or having a carbon-rich composition \citep{Madhusudhan2012,Miozzi2018}. We understand that the situation concerning 55 Cancri e is complex as there is limited observational data that may appear contradictory in some occasions. Nonetheless, here we propose a model that could potentially explain how some hot super-Earth can host atmospheres with large hydrogen components for timescales comparable to the lifespan of their host-stars.
	
	\item\textit{We assumed hydrogen does not chemically interact with the volcanic molecules in the diffusive phase.} We would expect some hydrogen to react with other species and become locked up therefore resulting in a diminished XUV-induced atmospheric erosion. Consequently our model overestimates the amount of available hydrogen which implies that the presence of an atmosphere on 55 Cancri e is even more problematic in the absence of a counter-mechanism.
	
	\item\textit{For Equation~\ref{eq:masslosstimescale} and \ref{eq:minmasstotal} we assumed the host star’s X-Ray and UV flux remain constant with time.} This approximation is appropriate for exoplanets orbiting G-type stars that become tidally locked within their first $\sim 50$ $\rm Myr$ since X-Ray and UV radiation is somewhat uniform. However, beyond this time our model gives an overestimation for the minimum mass required for a super-Earth to have a molecular hydrogen atmosphere. This however can easily be resolved by implementing the time-dependence accounted for in \citet{Penz2008(2),Sanz2011} and integrating with respect to time.
	
	\item\textit{The dayside has no atmosphere so it is modelled as a vacuum (Equations~\ref{eq:gasforce}).} The dayside is certainly not a vacuum; it instead might host a low-altitude atmosphere composed of volcanic and vaporised surface materials \citep[e.g.][]{Schaefer2009,Schaefer2012,Ito2015}. We assumed a vacuum in order to consider the maximum possible gas force and whether it could be balanced by the tidal and centrifugal forces. Since we were able to balance out the forces even in the extreme case, we argue that the tidal and centrifugal forces can also counteract the gas force for more realistic atmospheres.
	
	\item\textit{In Equation~\ref{eq:verbunt} we assume that $\sim 50\%$ of the angular momentum from the mass lost is conserved.} The true amount of angular momentum conserved is not easily calculated and would most probably require a fully hydrodynamic treatment. For instance, if $0\%$ of all angular momentum were conserved then according to \citet{Verbunt1993} the final orbital distance would be almost identical to the initial one. Therefore the initial mass of 55 Cancri e would be set by the amount of XUV it received during its lifetime. \citet{Locci2019} performed a backwards reconstruction of 55 Cancri e using this assumption and arrived at an initial mass of $\rm \sim 110 M_{\oplus}$. Conversely, if $100\%$ of all angular momentum were conserved then the exponent in equation~\ref{eq:verbunt} would change from $7/4 \Rightarrow 2$. This change would decrease the maximum initial mass (Equation~\ref{eq:maxmass}) and increase the minimum initial mass required for 55 Cancri e to have a hydrogen atmosphere (Equation~\ref{eq:minmass}). In other words, when more momentum is conserved the required initial conditions have to be more fine-tuned in order for our proposed mechanism to take effect. We assumed $\sim 50\%$ of the angular momentum was conserved as it is the middle point.
	
	\item\textit{We did not account for the effect of mass lost on the planetary rotation.} This is also a complicated hydrodynamical problem as it depends on how much of the planet's angular momentum is dissipated into the evaporating gases. In addition, other counter-mechanisms need to be considered for such as how friction caused by a large primordial atmosphere could slow down a planet's rotation. For instance, as an atmosphere is evaporated it becomes thinner so whilst on the one hand angular momentum could be extracted from the interior, on the other hand the loss of gas might result in less atmospheric friction. The balance of these mechanisms requires a full analysis in order to accurately determine the effects of the atmosphere on the planet's rotation. In any case, since we did not account for these effects we overestimated how long it would take for the planet to become tidally locked meaning that in reality less strict initial conditions would have been required for the planet to become tidally locked before losing the entirety of its primordial atmosphere.
	
	\item\textit{When calculating the maximum initial mass (Equation~\ref{eq:maxmassderive}), we assumed that the initial density of the exoplanet was equal to its host star’s one.} Our assumption is based on the observation that planets with large hydrogen envelops such as the outer planets in our solar system (except Saturn) have densities very similar to our Sun.
	
	\item\textit{In the case of 55 Cancri e we did not account for planet-planet interactions and how they could induce large scale migration.} \citet{Hansen2015} point out that in multiplanetary systems where the stellar-induced tidal evolution is coupled to the long-term non-periodic pertubations caused by the other planets, the inner planet can experience a strong semi-major axial evolution. They argue that the eccentricities and inclinations of 55 Cancri b and c changed in order to compensate for the significant migration experienced by 55 Cancri e. If this is the case, the maximum theoretical mass predicted by Equation~\ref{eq:maxmass} may be an underestimate and the true value may be significantly greater meaning that the initial mass can be constrained less thoroughly. Notwithstanding, as shown in figure~\ref{fig:55cancrie} a higher maximum initial mass increases the range of permissible initial rotational periods for the tidal locking timescale to be shorter than the mass-loss timescale. This raises the probability of an exoplanet having the right initial conditions for it to retain hydrogen in its atmosphere.
	
	\item\textit{In our simple analytic model we did not account for magnetic fields.} The effects of magnetism on atmospheric erosion is an active area of research where it is generally accepted that the presence of a B-field can slow down atmospheric destruction. For instance, planets are mostly well-shielded from coronal mass ejections, even for relatively weak intrinsic fields \citep[e.g.][]{Cohen2011}. Whilst the magnetic field of 55 Cancri e is unknown, using the dynamo scaling law from \citet{Christensen2006} with the core radius estimate from \citet{Crida2018} and the radiogenic models from \citet{Frank2014}, it can shown that a B-field of $\rm \sim B_{Earth}$ or potentially higher is reasonable. At this strength, the ionised hydrogen atoms in the outer layers of the atmosphere would be strongly influenced and mass-loss would be mitigated. However, even in the absence of a geomagnetic field, XUV irradiation should result in an ionosphere that would interact with escaping ions. A similar mechanism has been observed on Venus \citep{Zhang2012}. In conjunction, this implies that XUV atmospheric erosion would be more inefficient when a magnetic field is present. Therefore, a larger range of initial rotational periods would be permitted in order for condition~\ref{eq:minmasscondition} to hold. Consequently, including a magnetic field would increase the probability of super-Earths and sub-Neptunes having the right conditions for our mechanism to take place.
	
\end{enumerate}

\subsection{The Implications of Our Study}

\citet{Fulton2017} showed that there is a bimodal distribution in exoplanet radii with one peak at $\rm \sim 1.3 R_{\oplus}$, the other at $\rm \sim 2.4 R_{\oplus}$, and the minima at $\rm \sim 1.8 R_{\oplus}$. They argue that instead of using arbitrary definitions for super-Earths and sub-Neptunes, these should be defined more objectively. Specifically, super-Earths should be defined as having radii between $\rm 1 - 1.75R_{\oplus}$, while sub-Neptunes should have radii ranging from $\rm 1.75 - 3.5R_{\oplus}$. One of the most thorough analyses of the bimodal distribution in exoplanet sizes came from \citet{Owen2013, Owen2017}. In their model they argued that XUV-induced mass loss results in the observed $\rm \sim1.8 R_{\oplus}$ minima as Neptune-mass objects closer than $\sim 0.1$ AU lose their hydrogen envelops. As a general rule of thumb we agree with the conclusions derived in the aforementioned study. Our model just states that in some cases small and hot exoplanets may be able to retain atmospheres on their nightsides if they have the right configuration. It may be the case that the presence or lack of a hydrogen component in an atmosphere on a hot super-Earth could give information on its initial mass and day-length. Constraining the birth conditions of an exoplanet should provide more information on its planetary system's history, and potentially improve our overall understanding of exoplanetary statistics.

\section{Conclusions}

55 Cancri e may be an exoplanet with a configuration such that its nightside hosts an envelop with a considerable hydrogen component whilst its dayside perhaps sustains a volcanic and vaporised mineral atmosphere. For such a system to exist it is required that the thermal, gravitational, and centrifugal forces experienced by the nightside to balance out the gas force. Moreover, the exoplanet must become tidally locked before it has lost the entirety of its hydrogen atmosphere. Super-Earths/sub-Neptunes and their environments are complex systems so we are aware that we most probably did not account for all opinions, models and observations. In spite of this, our model was constructed from first principles and used data from peer-reviewed publications that cover a wide range of opinions, methodologies and conclusions from various fields such as astrophysics, geochemistry, geology and geophysics. With the information available to us at the time we believe we may have identified a mechanism through which some super-Earths/sub-Neptunes could retain hydrogen in their envelops. One potential method of testing our model would be to investigate the scale heights of the dayside and nightside of 55 Cancri e to see if there is a noticeable difference. A strong difference in the scale heights, once temperature is accounted for, would imply a difference in composition between the two faces which would support our hypothesis. With the \textit{James Webb Space Telescope (JWST)} expected to launch in 2021, the \textit{Atmospheric Remote-sensing Infrared Exoplanet Large-survey (ARIEL)} in 2028, the \textit{Thirty Meter Telescope (TMT)} and \textit{Extremely Large Telescope (ELT)} which have estimated first light times of 2024 and 2027 respectively; far superior astronomical observations would be possible. Consequently, uncertainties would be minimised hence allowing astronomers to thoroughly test our model and judge its validity.

\section*{Acknowledgements}

We acknowledge the support of the \emph{ARIEL ASI-INAF agreement n. 2018-22-HH.0} and the \emph{ERC grant ExoLights GA 617119}. This project has received funding from the European Union’s Horizon 2020 research and innovation programme 776403, ExoplANETS A  and the Science and Technology Funding Council (STFC) grants: ST/K502406/1 and ST/P000282/1. We thank Ignazio Pillitteri for useful discussions. We are grateful for the anonymous referee's valuable suggestions.

\bibliography{bibliography.bib}
\bibliographystyle{aasjournal}

\appendix

\section{Full Derivation of the Minimum Mass}

\begin{subequations}
	\begin{equation}
	t_{\rm mass\:loss} > t_{\rm lock}
	\end{equation}
	Substitute in Equation~\ref{eq:tidallocking} and \ref{eq:masslosstimescale}
	\begin{equation}
	\dfrac{G K M_{0} (M_{0} - M_{P})}{225 \varepsilon \pi R^{3}_{0} F_{\rm XUV}} > \dfrac{\omega_{0} a^{6}_{0} I_{0} Q_{0}}{3 G M^{2}_{0} k_{2} R^{5}_{0}}
	\end{equation}
	For super-Earths/sub-Neptunes with hydrogen envelops $ I_{0} \approx 0.3 M_{0} R^{2}_{0} $
	\begin{equation}
	\dfrac{G K M_{0} (M_{0} - M_{P})}{225 \varepsilon \pi R^{3}_{0} F_{\rm XUV}} > \dfrac{0.3\omega_{0} a^{6}_{0} Q_{0}}{3 G M_{0} k_{2} R^{3}_{0}}
	\end{equation}
	The flux, $F_{\rm XUV}$, can be substituted by $I_{\rm XUV}/4 \pi a^{2}_{0}$ and then the equation can be simplified.
	\begin{equation}
	\dfrac{8 G K M_{0} (M_{0} - M_{P})}{45 \varepsilon I_{\rm XUV}} > \dfrac{\omega_{0} a^{4}_{0} Q_{0}}{G M_{0} k_{2}}
	\end{equation}
	We substitute in Equation~\ref{eq:verbunt}
	\begin{equation}
	\dfrac{8 G K M_{0} (M_{0} - M_{P})}{45 \varepsilon I_{\rm XUV}} > \dfrac{\omega_{0} a^{4}_{P} Q_{0} M^{7}_{P}}{G M^{8}_{0} k_{2}}
	\end{equation}
	We rearrange the equation as follows
	\begin{equation}
	M^{10}_{0} \left(1 - \dfrac{M_{P}}{M_{0}} \right)  > \dfrac{45 \varepsilon I_{\rm XUV} Q_{0}}{8 G^{2} K k_{2}} \cdot \omega_{0} \cdot M^{7}_{P} \cdot a^{4}_{P}
	\end{equation}
	We raise everything to the ${1/10}^{\rm th}$ power and approximate the left-hand side (LHS) with the Taylor series
	\begin{equation}
	M_{0}\left(1 - \dfrac{M_{P}}{10M_{0}} \right) \gtrsim \left( \dfrac{45 \varepsilon I_{\rm XUV} Q_{0}}{8 G^{2} K k_{2}} \right)^{1/10} \cdot \omega^{1/10}_{0} \cdot M^{7/10}_{P} \cdot a^{2/5}_{P}
	\end{equation}
	Substituting in $ \omega_{0} = |{2 \pi}/{T_{orb}} - {2 \pi}/{\Delta T}| $ we find:
	\begin{equation}
	M_{0} \gtrsim \left( \alpha \cdot \left|  \dfrac{1}{T_{\rm orb}} - \dfrac{1}{\Delta T} \right| ^{1/10} \cdot M^{7/10}_{P} \cdot a^{2/5}_{P} \right)  + \dfrac{M_{P}}{10} 
	\end{equation}
\end{subequations}
\\
Where
$ \alpha =\left( \dfrac{45 \pi \varepsilon I_{XUV} Q_{0}}{4 G^{2} K k_{2}} \right)^{1/10} $



\end{document}